\documentclass[aps,reprint,amsmath]{revtex4-1}
\usepackage{graphicx}
\usepackage{dsfont}

\newcommand{\bra}[1]{\langle #1 | \,}
\newcommand{\ket}[1]{\, | #1 \rangle}
\newcommand{\expv}[1]{\langle #1 \rangle}
\newcommand{\ga}{\gamma}
\newcommand{\Ga}{\Gamma}

\newcommand{\de}{\delta}
\newcommand{\De}{\Delta}

\newcommand{\Om}{\Omega}
\newcommand{\hlf}{\mbox{$\frac{1}{2}$}}

\begin{document}

\title{Grover search algorithm with Rydberg-blockaded atoms: \\
Quantum Monte Carlo simulations}

\author{David Petrosyan}
\affiliation{Institute of Electronic Structure and Laser, FORTH,
GR-71110 Heraklion, Crete, Greece}

\author{Mark Saffman}
\affiliation{Department of Physics, University of Wisconsin-Madison,
Madison, Wisconsin 53706, USA}

\author{Klaus M\o lmer}
\affiliation{Department of Physics and Astronomy, Aarhus University,
DK-8000 Aarhus C, Denmark}

\date{\today}

\begin{abstract}
We consider the Grover search algorithm implementation for a quantum
register of size $N=2^k$ using $k$ (or $k+1$) microwave- and laser-driven
Rydberg-blockaded atoms, following the proposal by
M\o lmer, Isenhower, and Saffman [J. Phys. B \textbf{44}, 184016 (2011)].
We suggest some simplifications for the microwave and laser couplings,
and analyze the performance of the algorithm for up to $k=4$
multilevel atoms under realistic experimental conditions using
quantum stochastic (Monte-Carlo) wavefunction simulations.
\end{abstract}


\maketitle

\section{Introduction}

Strong, long-range interactions between atoms in high-lying Rydberg
states make them attractive systems for quantum information
applications \cite{rydQIrev}.
The interaction-induced level shifts suppress resonant optical
excitation of Rydberg states of more than one atom within a
certain blockade distance from each other \cite{rydQIrev,Comparat2010}.
This blockade effect can then be used to implement quantum logic gate
operations between closely spaced atoms
\cite{Jaksch2000,LIMSKM2011,NatPRLSaffman,NatPRLGrangier,Beguin2013,Maller2015},
or to realize atomic ensemble qubits with Rydberg superatoms
which can accommodate at most one collective Rydberg excitation
at a time \cite{Lukin2001,Kuzmich2012,Weber2015,Ebert2015,Zeiher2015}.

The Grover quantum search algorithm \cite{Grover1997}, which
offers a quadratic speed-up of the search of unstructured
databases over classical search algorithms, is a paradigmatic
example of the power of quantum computation \cite{MNICh2000,PLDP2007}.
The protocol consists of repeated application of the query (oracle)
and inversion-about-the-mean (Grover) operations to a quantum register
composed of $k$ qubits which can store $2^k$ elements (database entries).

As any other quantum computation procedure, both the oracle and
Grover operations can be implemented by a sequence of standard,
universal one- and two-qubit gates \cite{MNICh2000}.
When the number of qubits $k$ increases beyond just a few,
however, such implementations become quite complex
and experimental demonstrations of the Grover search
algorithm have so far been restricted to the case of $k=2$
\cite{Chuang98, Brickman05, Walther05, Schoelkopf09}.

In contrast, an efficient procedure to implement the Grover search
algorithm using the multi-atom interactions mediated by the Rydberg
blockade was proposed in Ref. \cite{KMLIMS2011}. In that proposal,
individual qubits are encoded in pairs of metastable states of single
atoms trapped in an array of microtraps, and the oracle and Grover
operations require only simple sequences of excitation and deexcitation
processes between the qubit states and the Rydberg state in each atom.
Here we suggest a practical implementation of this proposal with
microwave and laser drivings of the atoms. We perform extensive
numerical simulations of the dynamics of the system under realistic
assumptions about the interatomic interaction strengths
as well as atomic decay and dephasing parameters.
We present results for the success probabilities of the Grover
search with a moderate -- but computationally nontrivial -- register
size of $k \leq 4$. We explore two different interaction configurations
proposed in Ref. \cite{KMLIMS2011}, wherein the blockade interaction
is present either between any pair of register atoms excited to the
Rydberg state, or only between an auxiliary atom and each register
atom. Both configurations have advantages and disadvantages for
the experimental realization, and we find that they yield similar
performance of the algorithm.

\section{The atomic system}

\begin{figure}[t]
\includegraphics[width=8.7cm]{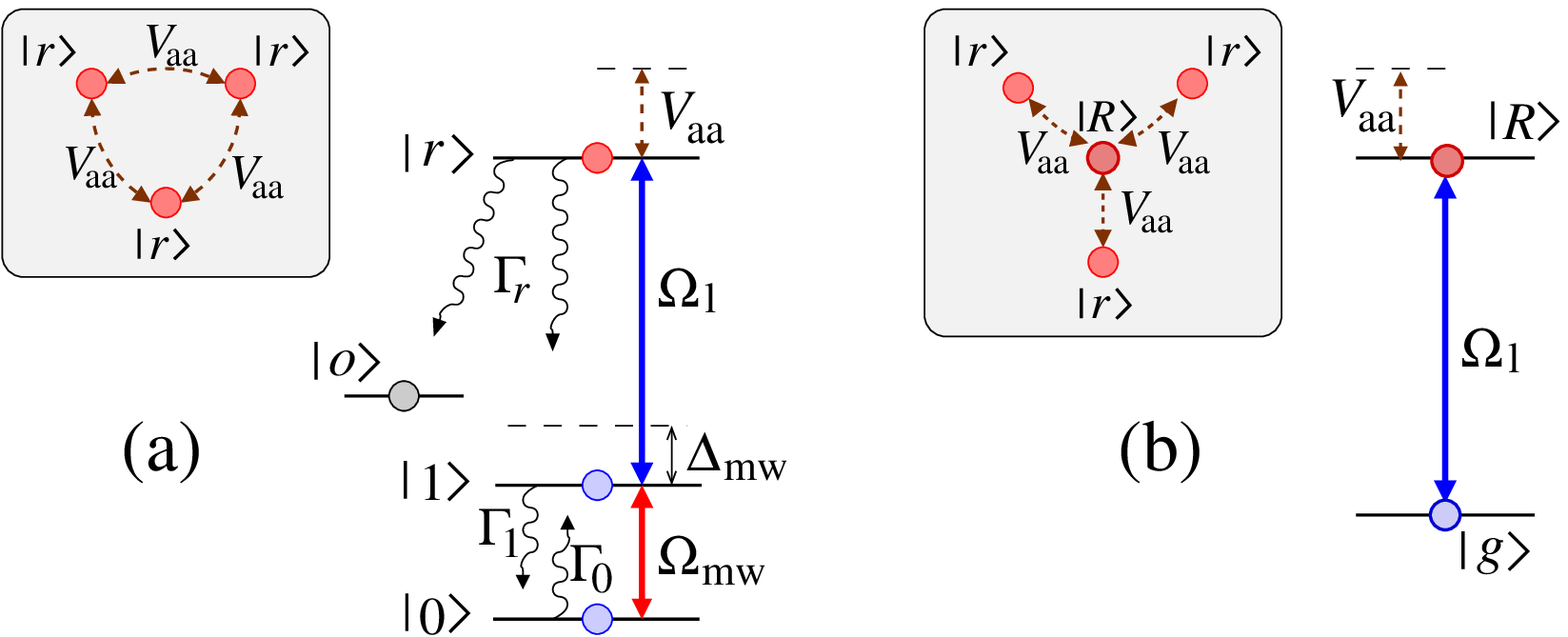}
\caption{(a) Level scheme of the register atoms interacting with
a microwave field on the transition $\ket{0} \leftrightarrow \ket{1}$
with Rabi frequency $\Om_{\mathrm{mw}}$, and with resonant laser
field(s) on the transition $\ket{1} \leftrightarrow \ket{r}$
with Rabi frequency $\Om_{\mathrm{l}}$.
The coupling of selected atoms with the global microwave field
can be made resonant ($\De_{\textrm{mw}} \ll |\Om_{\mathrm{mw}}|$) or
strongly detuned ($\De_{\textrm{mw}} \gg |\Om_{\mathrm{mw}}|$).
State $\ket{o}$ accounts for the loss of the atom due to decay
from $\ket{r}$ to any other state but $\ket{0}$ or $\ket{1}$.
Inset: Atoms in Rydberg states $\ket{r}$ interact with each other
via a strong, long-range potential $V_{\mathrm{aa}} \gg |\Om_{\mathrm{l}}|$
which suppresses Rydberg excitation of all but one atom at a time.
(b) Level scheme of an ancilla atom whose transition
$\ket{g} \leftrightarrow \ket{R}$ is driven by a focused resonant
laser with Rabi frequency $\Om_{\mathrm{l}}$. Inset: The ancilla atom
in Rydberg state $\ket{R}$ interacts with all the $\ket{r}$-state
register atoms via the strong potential $V_{\mathrm{aa}}$, while
the register atoms do not directly interact with each other.}
\label{fig:als}
\end{figure}

Consider $k$ atoms with the level scheme sketched in Fig.~\ref{fig:als}(a).
States $\ket{0}$ and $\ket{1}$ of each atom are the qubit basis
states. A time-dependent microwave field acts on the transition
$\ket{0} \leftrightarrow \ket{1}$ with the Rabi frequency
$\Om_{\mathrm{mw}}(t) = |\Om_{\mathrm{mw}}| e^{i \phi}$ having real
amplitude $|\Om_{\mathrm{mw}}|$ and phase $\phi$.
The corresponding Hamiltonian for the $j$th atom is
\begin{equation}
H_{\textrm{mw}}^{(j)} = -\hbar [\De_{\textrm{mw}}^{(j)} \sigma_{11}^{(j)} +
\hlf \Om_{\textrm{mw}} \sigma_{10}^{(j)} + \hlf \Om_{\textrm{mw}}^* \sigma_{01}^{(j)}] ,
\end{equation}
where $\sigma_{\mu \nu}^{(j)} \equiv \ket{\mu}_j \bra{\nu}$ are the atomic
operators, and $\De_{\textrm{mw}}^{(j)}$ is the microwave field detuning.

A resonant field, $\De_{\textrm{mw}} = 0$, applied to the atom leads
to the unitary transformation (in the qubit basis $\{\ket{0},\ket{1}\}$)
\begin{equation}
U_{\phi}(\theta) =  \left[ \begin{array}{cc}
\cos \hlf \theta & i e^{-i \phi} \sin \hlf \theta \\
i e^{i \phi} \sin \hlf \theta & \cos \hlf \theta
\end{array} \right] \; , \label{eq:qU}
\end{equation}
where $\theta = \int |\Om| dt$ is the pulse area \cite{PLDP2007}.
Hence, $\theta = \pi$ with $\phi = 0 \, (\pi/2)$ corresponds to the
$iX \, (iY)$ operation on the qubit \cite{MNICh2000}. The $Z$ gate
can be realized as $U_{\pi/2}(\pi) U_{0}(\pi) = iZ$, which, for a fixed
maximum amplitude of $|\Om|$, takes twice the time of the $X$ or $Y$ gate.
The Hadamard gate can be realized as $U_{\pi/2}(\pi/2) U_{0}(\pi) = iH$,
which takes 1.5 times longer than $X$ or $Y$. In what follows,
we will only use the operations $U_{0}(\pi) = iX$ and the Hadamard-like
$U_{\pm \pi/2}(\pi/2)$ which takes only half the time of $X$.

Since the distance between the atoms is small --- of the order of a few $\mu$m
\cite{NatPRLSaffman,NatPRLGrangier,Beguin2013,Maller2015}
for the Rydberg blockade to be effective (see below) ---
the microwave field with a long wavelength of several cm
affects all the atoms with the same Rabi frequency. We assume that
the frequency of the microwave field can be tuned into, or detuned from,
the transition resonance $\ket{0} \leftrightarrow \ket{1}$ of the unperturbed
atoms. In addition, we assume that using focused, non-resonant laser
beams we can induce ac Stark-shifts of, e.g., state $\ket{1}$ of
the selected atoms to make the transition $\ket{0} \leftrightarrow \ket{1}$
resonant with the microwave field (when it is non-resonant otherwise),
or to detune it by a large amount $\De_{\textrm{mw}} \gg |\Om_{\mathrm{mw}}|$
(when it is resonant otherwise) \cite{Xia2015}. We can therefore
selectively couple or decouple the atoms to and from the global
microwave field.

Each register atom $j$ in state $\ket{1}$ can be excited by a focused
laser beam to a Rydberg state $\ket{r}$, see Fig.~\ref{fig:als}(a).
This process is described by the Hamiltonian
\begin{equation}
H_{\textrm{l}}^{(j)} = - \hbar \hlf [\Om_{\textrm{l}}^{(j)} \sigma_{r1}^{(j)}
+ \Om_{\textrm{l}}^{*(j)} \sigma_{1r}^{(j)} ] ,
\end{equation}
were $\Om_{\textrm{l}}^{(j)}$ is the Rabi frequency, and we assume that the
laser is resonant for an unperturbed (not blockaded) atom, leading to
the same transformations as in Eq.~(\ref{eq:qU}) between states
$\ket{1}$ and $\ket{r}$. We will also employ an auxiliary (ancilla) atom $a$
with levels $\ket{g}$ and $\ket{R}$ similarly coupled by a focused resonant
laser, as shown in Fig.~\ref{fig:als}(b).

As in Ref. \cite{KMLIMS2011}, we will consider two possible
scenarios of interatomic interactions.
In the first case [see the inset of Fig.~\ref{fig:als}(a)],
any pair of register atoms $i$ and $j$ in state $\ket{r}$ interact
with each other via the long-range potential
\begin{equation}
H_{\textrm{aa}}^{(i,j)} = \hbar \frac{C_p}{|\mathbf{r}_i - \mathbf{r}_j|^p}
\sigma_{rr}^{(i)} \otimes \sigma_{rr}^{(j)} , \label{eq:Hintrr}
\end{equation}
where $\mathbf{r}_{i,j}$ are the atomic positions, and $p=3$ or $6$
for the dipole-dipole or van der Waals interactions, respectively
\cite{rydQIrev}. If the interaction-induced level shifts
$V_{\textrm{aa}} = C_p/r_{ij}^p$ are large enough, then an atom
already excited to the Rydberg state $\ket{r}$ will block
subsequent excitation of all the other atoms.
In the second case [see the inset of Fig.~\ref{fig:als}(b)],
we assume that the register atoms do not interact with each other,
but each register atom $j$ in state $\ket{r}$ interacts with
the ancilla atom $a$ in state $\ket{R}$ via the potential
\begin{equation}
H_{\textrm{aa}}^{(j,a)} = \hbar \frac{C_p}{|\mathbf{r}_j - \mathbf{r}_a|^p}
\sigma_{rr}^{(j)} \otimes \sigma_{RR}^{(a)} , \label{eq:Hintar}
\end{equation}
which can block the Rydberg excitation of the ancilla atom in
the presence of one or more $\ket{r}$-state register atoms.
This situation occurs for example for certain configurations
of Rydberg excited states in rubidium and cesium \cite{Beterov2015}.

We shall include in our treatment realistic atomic decay and
dephasing, leading to decoherence and loss of atoms which
strongly affect the outcome of the quantum computation.
The atoms are subject to the following relaxation processes:
slow decays of level $\ket{1}$ to $\ket{0}$ with rate $\Ga_1$
and level $\ket{0}$ to $\ket{1}$ with rate $\Ga_0$ \cite{Xia2015};
the much faster decay of the Rydberg state $\ket{r}$ with rate
$\Ga_r = \Ga_{r0} +  \Ga_{r1} + \Ga_{ro}$ which has three contributions:
the decay to $\ket{0}$, to $\ket{1}$ and loss of population to
any other state represented in our model by $\ket{o}$ \cite{Maller2015};
finally, we include dephasing $\ga_{z}$ on the qubit microwave
transition $\ket{0} \leftrightarrow \ket{1}$ and the typically
much stronger dephasing $\ga_{r}$ of the atomic polarization
on the optical transition $\ket{1} \leftrightarrow \ket{r}$ due to,
e.g., the laser phase fluctuations, external field noise, and residual
decay to $\ket{1}$ via a non-resonant intermediate excited state
$\ket{e}$ (relevant, when $\ket{1} \leftrightarrow \ket{r}$ is
a two-photon transition via $\ket{e}$).
The corresponding Lindblad generators \cite{MNICh2000,PLDP2007}
for the decay and dephasing processes are
$L_{01}^{(j)} = \sqrt{\Ga_{0}} \sigma_{10}^{(j)}$,
$L_{10}^{(j)} = \sqrt{\Ga_{1}} \sigma_{01}^{(j)}$,
$L_{r0}^{(j)} = \sqrt{\Ga_{r0}} \sigma_{0r}^{(j)}$,
$L_{r1}^{(j)} = \sqrt{\Ga_{r1}} \sigma_{1r}^{(j)}$,
$L_{ro}^{(j)} = \sqrt{\Ga_{ro}} \sigma_{or}^{(j)}$, and
$L_{\mathrm{mw}}^{(j)} = \sqrt{\ga_{z}/2} (2 \sigma_{11}^{(j)} - \mathds{1}^{(j)})$,
$L_{\mathrm{opt}}^{(j)} = \sqrt{\ga_{r}/2} (2 \sigma_{rr}^{(j)} - \mathds{1}^{(j)})$,
where $\mathds{1}^{(j)} \equiv \sum_{\mu} \sigma_{\mu\mu}^{(j)}$ is the unity
operator for atom $j$.

For an isolated atom, the excitation linewidth of the Rydberg
state $\ket{r}$ (from state $\ket{1}$) is 
$w = |\Om_{\textrm{l}}| \sqrt{\ga_{r1}/\Ga_r}$,
where $\ga_{r1} \equiv \frac{1}{2} (\Ga_1 + \Ga_r) + \ga_{r}$ and
$|\Om_{\textrm{l}}|^2 \gg  \Ga_r \ga_{r1}$ \cite{PLDP2007}.
In what follows, we position the atoms such that the interaction-induced
level shifts $V_{\textrm{aa}} \geq 10 w$ are sufficiently large
for the Rydberg blockade of any pair of register atoms $i,j$,
or any register atom $j$ and ancilla atom $a$.

\section{The search algorithm implementation}

With the Grover algorithm, we search for one particular marked element
$x_m = b_0 b_1 \cdots b_{k-1}$ ($b_j \in [0,1]$) in a database containing
$N=2^k$ elements $x=00 \cdots 0,  \; 00 \cdots 1, \; \ldots , \; 11 \cdots 1$.
The algorithm consists of the following steps
\cite{Grover1997,MNICh2000,PLDP2007}:

\begin{itemize}

\item[0)] prepare the $k$-qubit register in an equally-weighted
superposition $\ket{s} \equiv
\left[ \frac{\ket{0} + \ket{1}}{\sqrt{2}} \right] ^{\otimes k}
= \frac{1}{\sqrt{N}} \sum_{x} \ket{x}$ of all $N$ possible states $\ket{x}$;

\item[1)] apply to the register the oracle query operation
which shifts the phase of state $\ket{x_m} = \ket{b_0 b_1 \cdots b_{k-1}}$
by $\pi$ (flips the sign of $c_{x_m}$) relative to all the other
states $\ket{x}$ of the superposition $\sum_{x=00\ldots0}^{11\ldots1} c_x \ket{x}$;

\item[2)] apply to the register the inversion about the mean (Grover)
operation.

\end{itemize}

The register preparation step 0) is applied only once. 
The combined effect of steps 1) and 2) is to increase 
the amplitude $c_{x_m}$ of state $\ket{x_m}$ by $\sim 1/\sqrt{N}$ 
at the expense of amplitudes $c_{x}$ of all the other states $\ket{x}$. 
Steps 1) and 2) are thus applied repeatedly, $\sim \sqrt{N}$ times, 
until the probability of the marked state approaches unity, at which 
time it is measured.

We now examine in some detail the implementation of each of the above
steps, along the lines of the proposal of Ref.~\cite{KMLIMS2011}
with a view of possible experimental realization \cite{Maller2015,Xia2015}.

0) The register preparation step in Ref.~\cite{KMLIMS2011} is performed
in the standard way by applying the Hadamard gate $H$ to all the
register atoms initially in state $\ket{0}$. The resonant microwave
implementation of $H$ would involve $\pi+\pi/2$ pulses which take
1.5 times the duration of the $X$ gate, but the initial superposition
state can also be obtained by the shorter transformation $U_{-\pi/2}(\pi/2)$
applied \textit{simultaneously} to all the atoms (qubits) in state $\ket{0}$:
$U_{-\pi/2}(\pi/2) \ket{0} \to [\ket{0} + \ket{1}]/\sqrt{2}$.

1) In the oracle step, the protocol of Ref.~\cite{KMLIMS2011} applies
sequentially to each register atom $j=0,1,\ldots,k-1$ a $\pi$-pulse $U_{0}(\pi)$
between states $\ket{1-b_j}$ and $\ket{r}$ using focused laser beams;
if any one atom is transferred to $\ket{r}$, then all the subsequent
atoms remain in their initial state $\ket{0}$ or $\ket{1}$ due to the
Rydberg blockade [assuming the interaction scenario of Eq.~(\ref{eq:Hintrr})].
This is then followed by the same operation $U_{0}(\pi)$ on all the register
atoms in the reverse order to bring the atom in Rydberg state $\ket{r}$
back to its initial state. The result of this transformation is that
any state of the register $\ket{\mu_0 \mu_1 \cdots \mu_{k-1}}$ ($\mu =0,1$)
having one or more digits different from the marked state
$\ket{b_0 b_1 \cdots b_{k-1}}$ will undergo one (and not more than one,
due to the Rydberg blockade) full Rabi cycle via the Rydberg state
and accumulate a $\pi$ phase shift, while only the marked state
$\ket{b_0 b_1 \cdots b_{k-1}}$ will remain unchanged.

We now assume that the Rydberg exciting lasers with fixed frequency
act only on the transition $\ket{1} \to \ket{r}$. We should therefore
implement the oracle step by applying first the $b_i X$ transformation
to each register atom, which flips the qubit states $\ket{0}$ and $\ket{1}$
if $b_i = 1$ and does nothing otherwise, and then apply the Rydberg excitation
and de-excitation lasers, followed again by the $b_i X$.
To implement the $b_i X$ with our setup, we apply the global microwave
$U_{0}(\pi) = iX$ pulses to all the atoms \textit{simultaneously}, but
we use Stark lasers to adjust each atom's detuning with respect to the
microwave frequency to $(1-b_i) \De_{\textrm{mw}}$, respectively \cite{Xia2015}.

2) In the Grover step, Ref.~\cite{KMLIMS2011} proposes to transfer
sequentially each register atom in state $[\ket{0} - \ket{1}]/\sqrt{2}$
to the Rydberg state $\ket{r}$ and then back in reverse order, while
leaving the atoms in the ``dark'' state $[\ket{0} + \ket{1}]/\sqrt{2}$
unaffected by the Rydberg lasers. Again, Rydberg excitation of any
one atom would block subsequent excitation of the other atoms
[assuming the interactions of Eq.~(\ref{eq:Hintrr})].
This transformation leaves the equally-weighted superposition state
$\ket{s}$ unchanged, while flipping the sign of all the other orthogonal
states of the register, which results in the inversion about the mean
operation \cite{Grover1997,MNICh2000,PLDP2007}.

We implement this Grover operation by first applying to all
the atoms \textit{simultaneously} the microwave $U_{\pi/2}(\pi/2)$
pulse which results in transformation
\begin{eqnarray*}
U_{\pi/2}(\pi/2)[\ket{0} + \ket{1}]/\sqrt{2} &\to & \ket{0} \\
U_{\pi/2}(\pi/2)[\ket{0} - \ket{1}]/\sqrt{2} &\to &- \ket{1} .
\end{eqnarray*}
We then apply the resonant Rydberg excitation and de-excitation lasers
on the transition $\ket{1} \to \ket{r}$. Finally, we apply to all the
atoms \textit{simultaneously} the microwave $U_{-\pi/2}(\pi/2)$ pulse
which leads to
\begin{eqnarray*}
U_{-\pi/2}(\pi/2) \ket{0} &\to & [\ket{0} + \ket{1}]/\sqrt{2} \\
U_{-\pi/2}(\pi/2) (- \ket{1}) &\to &  [\ket{0} - \ket{1}]/\sqrt{2},
\end{eqnarray*}
as was required.

In both steps 1) and 2) above, the conditional logic operations
rely on the Rydberg blockade.
In the interaction scenario of Eq.~(\ref{eq:Hintrr}) (i.e., any pair
of register atoms in state $\ket{r}$ strongly interact with each other),
we apply the Rydberg excitation laser $\pi$-pulses $U_{0}(\pi)$ between
states $\ket{1}$ and $\ket{r}$ sequentially to atoms $j=0,1,\ldots,k-1$,
followed by the same de-excitation $\pi$-pulses $U_{0}(\pi)$ applied
to the atoms in the reverse order $j=k-1,k-2, \ldots,0$ \cite{KMLIMS2011}.
In the experimentally slightly simpler case of interaction of
Eq.~(\ref{eq:Hintar}) involving an ancilla atom (i.e., any register atom
in state $\ket{r}$ interacts only with the ancilla atom blocking its
excitation to state $\ket{R}$), we can apply the Rydberg excitation laser
$\pi$-pulse $U_{0}(\pi)$ to all the register atoms $j$ \textit{simultaneously},
transferring any atom in state $\ket{1}$ to state $\ket{r}$.
We then apply a $2\pi$-pulse $U_{0}(2\pi)$ to the ancilla atom
on the transition $\ket{g} \leftrightarrow \ket{R}$:
if one or more register atoms are in state $\ket{r}$, the ancilla
atom will remain in state $\ket{g}$ due to the Rydberg blockade;
and only if no register atom is in state $\ket{r}$, the ancilla atom
will undergo a full Rabi cycle between states $\ket{g}$ and $\ket{R}$
resulting in the sign change of the state of the combined system
consisting of the register atoms and the ancilla.
We then apply \textit{simultaneously} to all the register atoms
the de-excitation laser $\pi$-pulse $U_{\pi}(\pi)$ with the opposite
sign ($\phi_{\mathrm{l}} = \pi$ phase) of the Rabi frequency $\Om_{\mathrm{l}}$
so as to avoid the sign change of state $\ket{1}$.

\begin{figure}[t]
\includegraphics[width=8.7cm]{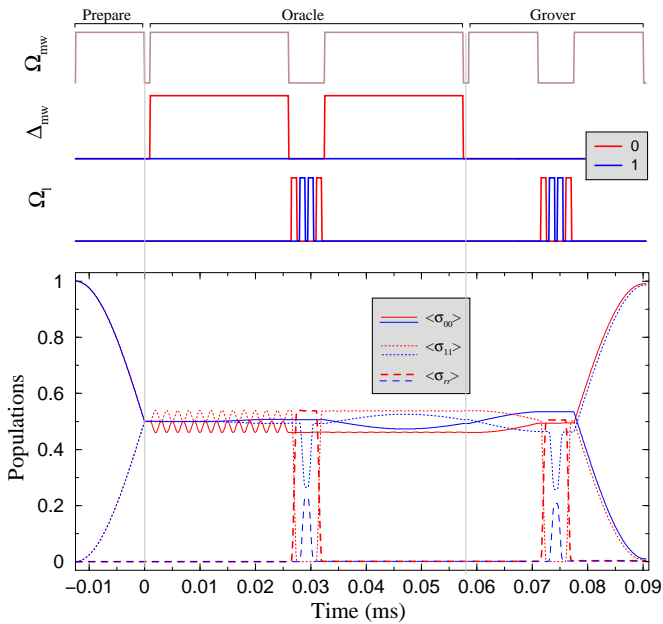}
\caption{Time-dependence of the microwave and laser fields (top)
acting on atoms $j=0,1$ (see the legend for color code),
and populations $\expv{\sigma_{\mu \mu}^{(j)}}$ of states
$\ket{\mu=0,1,r}$ of the corresponding atoms (main panel),
for one iteration of the search algorithm in a quantum register
of $k=2$ atoms interacting via Eq.~(\ref{eq:Hintrr}).
The marked input is $b_0b_1 =01$.}
\label{fig:algexe}
\end{figure}

\begin{figure}[t]
\includegraphics[width=8.7cm]{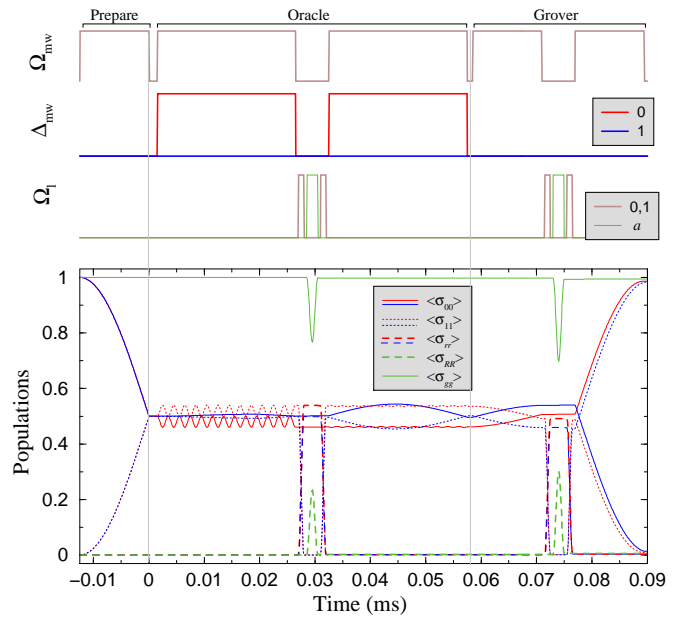}
\caption{Same as in Fig.~\ref{fig:algexe}, but with ancilla atom $a$
interacting with register atoms $j=0,1$ via Eq.~(\ref{eq:Hintar}).}
\label{fig:algexeX}
\end{figure}

To illustrate the foregoing discussion, in
Figs.~\ref{fig:algexe} and \ref{fig:algexeX} we plot
the time-dependence of the microwave and laser pulses and the
resulting coherent dynamics of populations of the atomic states.
In these figures, we show one full iteration of the search algorithm
with $k=2$ register atoms (plus the ancilla in Fig.~\ref{fig:algexeX})
and a representative marked input, assuming negligible relaxation rates.

\section{Results of simulations}

We simulate the dissipative dynamics of the system of $k$ atoms
using the quantum stochastic (Monte Carlo) wavefunctions method
\cite{PLDP2007,qjumps}. Using realistic experimental parameters,
we test various inputs $b_0 b_1 \cdots b_{k-1}$ and for each input
generate many independent trajectories for the time-evolution of
the wavefunction of the system.

We assume that after each run, the experimentalist performs a projective
measurement of all the register atoms (qubits) onto state $\ket{0}$.
Then the negative outcome of the measurement on some atom $j$
(i.e. the atom is not in state $\ket{0}$) would lead the experimentalist
to assume $b_j=1$ (and the atom is in state $\ket{1}$), but the same
measurement outcome would correspond also to atom $j$ being lost all
together (the atom is in state $\ket{o}$). Thus, if we average over
all possible inputs, a loss of an atom still leads to correct
measurement result half of the time. 
It is a special feature of our Rydberg blockade implementation of 
the search algorithm that if an atom is lost during the calculation, 
the oracle and Grover operations still apply correctly to the 
remaining string of qubits \cite{DasariLoss} 

\begin{figure}[t]
\hspace{-0.5cm}\includegraphics[width=8.7cm]{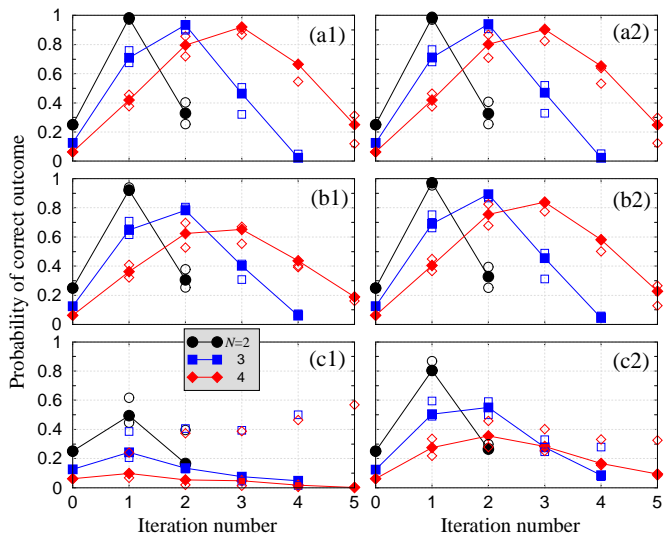}
\caption{Probabilities of measuring correct outcomes $b_0 b_1 \cdots b_{k-1}$
of the Grover search versus number of iterations, for $k=2,3,4$ digit
register (black, blue, red, respectively) with the interaction scheme
of Fig.~\ref{fig:als}(a) [Eq.~(\ref{eq:Hintrr})], obtained from averaging
over 200 independent trajectories for the system wavefunction.
The marked inputs for the filled symbols are $01, 010, 0101$;
other inputs, e.g., $00, 000, 0000$ and $11, 111, 1111$ shown
with open symbols, lead to close results to within $\pm 5\%$
for (a1,a2,b1,b2), or are more divergent for (c1,c2)
[typically, success probabilities for inputs $11,\ldots$ are
better than for $00,\ldots$, see the text for discussion].
The Rabi frequency of the Rydberg excitation laser is
$|\Om_{\mathrm{l}}| = 2 \pi \times 0.5\:$MHz in the left panels (a1,b1,c1) and
$|\Om_{\mathrm{l}}| = 2 \pi \times 2\:$MHz in the right panels (a2,b2,c2).
The Rydberg state decay is taken
$\Ga_r = (1,4.76,100) \times 10^3\:\mathrm{s}^{-1}$ in (a,b,c), respectively,
with $\Ga_{ro} = \frac{7}{8} \Ga_r$ and $\Ga_{r0},\Ga_{r1} = \frac{1}{16} \Ga_r$.
The dephasing rates on the Rydberg transitions are
$\ga_r = (1,10,100) \times 10^3 \:\mathrm{s}^{-1}$ in (a,b,c), respectively.
Other parameters, common to all the graphs, are
$\Ga_0,\Ga_1 = 2 \: \mathrm{s}^{-1}$, $\ga_z = 100 \: \mathrm{s}^{-1}$,
$|\Om_{\mathrm{mw}}| = 2 \pi \times 20\:$kHz
($X$ gate time is $25\:\mu$s), and $\De_{\textrm{mw}} = 25 |\Om_{\mathrm{mw}}|$,
while the time interval between the gates is $\de t = 50\:$ns.}
\label{fig:sucpritr}
\end{figure}

\begin{figure}[t]
\hspace{-0.5cm}\includegraphics[width=8.7cm]{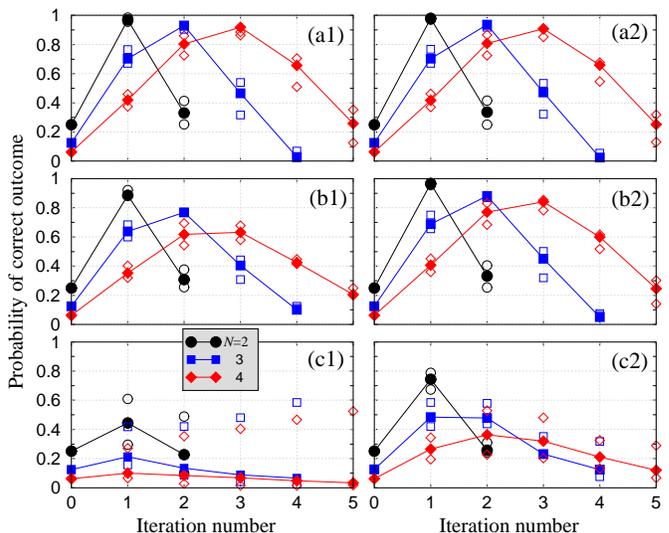}
\caption{Same as in Fig.~\ref{fig:sucpritr}, but for the interaction
scheme of Fig.~\ref{fig:als}(b) [Eq.~(\ref{eq:Hintar})].
Decay and dephasing of the ancilla atom are neglected.}
\label{fig:sucpritrX}
\end{figure}

In Figs.~\ref{fig:sucpritr} and \ref{fig:sucpritrX} we show the result
of our simulations for the interaction configurations of
Fig.~\ref{fig:als}(a) and (b) [Eqs.~(\ref{eq:Hintrr}) and (\ref{eq:Hintar})]
without and with the ancilla atom, respectively.
The probabilities of detecting the system in the correct marked
state $\ket{b_0 b_1 \cdots b_{k-1}}$ are obtained upon averaging over
many independent realizations (trajectories) of the numerical experiment.
More precisely, for an input of say $b_0 b_1 b_2 = 0 1 0$ we calculate
the probability of detecting the system in state
$\ket{\mu_0=0,\mu_1 \neq 0, \mu_2=0}$
($\mu \neq 0$ is either $\mu = 1$ or $\mu = o$).

In both Figs.~\ref{fig:sucpritr} and \ref{fig:sucpritrX} we obtain
similar results; only for very strong decay of the Rydberg state
corresponding to panels (c1,c2) the scheme with the ancilla atom
performed somewhat worse, even though we neglected the decay and dephasing
of the ancilla in Fig.~\ref{fig:sucpritrX} for a fair comparison with
Fig.~\ref{fig:sucpritr}. This is due to the fact that the scheme
with the ancilla permits multiple Rydberg excitations of the
register atoms, leading to their larger aggregate probability
of decay and loss.

We note that, for moderate values of the atomic decay and dephasing,
digits $b_j = 0$ in the marked element tend to cause larger error in the
outcome, because the microwave detuning $\De_{\textrm{mw}} = 25 |\Om_{\textrm{mw}}|$,
which suppresses the $X$ gate on atom $i$ during the oracle operation,
is large but still finite. More important are the relaxation processes
which significantly degrade the performance of the algorithm with
increasing evolution time. As a consequence, the probability for
the correct measurement outcome may peak after fewer iterations
than would be required to reach unity in the ideal case.
It turns out that the errors due to the decay and dephasing on
the qubit transition $\ket{0} \leftrightarrow \ket{1}$ play a minor role,
despite the slowness of the operations performed by the microwave field
with small Rabi frequency \cite{Xia2015}. The larger decay rate
$\Ga_r \simeq 5-100 \times 10^3\:\mathrm{s}^{-1}$ and higher probability
of atom loss from the Rydberg state \cite{Maller2015} are more damaging,
and the most harmful element is the large dephasing
$\ga_r = 10^5 \:\mathrm{s}^{-1}$ of Rydberg transition.
So either the laser Rabi frequency on the Rydberg transition should
be increased, as in panels (a2,b2,c2) of Figs.~\ref{fig:sucpritr}
and \ref{fig:sucpritrX}, so that the decay and dephasing have less
time to destroy the atomic coherences, or $\ga_r$ should be reduced.
While for the dephasing rate $\ga_r$ we took a typical experimental value,
there is no theoretical argument why this value could not be reduced
by an order of magnitude or more.

\section{Summary}

To conclude, we have studied the Grover search algorithm implementation
with several Rydberg blockaded atoms under realistic experimental
conditions including the choice of parameters for the atomic decay,
dephasing and interaction strengths.
We have shown that relaxation processes cause decoherence during
the quantum computation and reduce the probability of the correct
outcome after a few iterations of the oracle and Grover steps.

The remarkable property of the Grover algorithm is that it can
tolerate moderate amount of errors without error correction,
with the measurement on the final state of the system still
leading to increased probability of the sought-after element
of the database. When the probability for the correct outcome
is larger than all the probabilities for incorrect outcomes,
one may have recourse to perform several experimental runs
and measurements and obtain the correct result by a majority vote.


\begin{acknowledgments}
This work was supported by the US IARPA MQCO program,
the EU H2020 FET-Proactive project RySQ, and the Villum Foundation.
\end{acknowledgments}

\end{document}